\documentclass[aps,prd,10pt,
onecolumn,
showpacs,preprintnumbers,amsmath,amssymb,
a4paper,nofootinbib,longbibliography]{revtex4-2}

\usepackage{setspace}
\usepackage{centernot}
\usepackage{nicefrac}
\usepackage{empheq}
\usepackage{adjustbox}
\usepackage{fancybox}
\usepackage[most]{tcolorbox}
\usepackage{threeparttable}

\usepackage{esint}

\usepackage{empheq}
\usepackage{mathtools}

\usepackage{natbib}
\usepackage{amssymb,amsbsy,amsmath,amsfonts}
\usepackage{graphicx}
\usepackage{epsf,epsfig,float,latexsym,amsthm,fancyhdr,rotating}
\usepackage{graphics,psfrag,longtable}
\usepackage{slashed}
\usepackage[normalem]{ulem}
\usepackage[colorlinks,citecolor=blue,linktoc=all,linkcolor=cyan,urlcolor=blue]{hyperref}
\usepackage{upgreek}
\usepackage{siunitx}
\usepackage{dcolumn}

 \usepackage{multirow}
 \usepackage[compat=1.1.0]{tikz-feynman}
\usetikzlibrary{decorations.markings,arrows.meta}

\tikzset{
  myphoton/.style={decorate,
    decoration={snake, amplitude=1.5pt, segment length=5pt}},
  fsblob/.style={circle, fill=cyan!55, draw=cyan!70!black,
                 line width=0.7pt, minimum size=0.35cm, inner sep=0pt},
  hvpblob/.style={circle, fill=orange!15, draw=orange!80!black,
                  line width=0.8pt, minimum size=0.9cm, inner sep=0pt},
  vploop/.style={postaction={decorate, decoration={markings,
    mark=at position 0.75 with {\arrow[thick]{Stealth}}}}},
}

\def\beq{\begin{equation}}
\def\eeq{\end{equation}}
\def\bea{\begin{eqnarray}}
\def\eea{\end{eqnarray}}
\def\beqa{\begin{equation}\begin{array}{l}}
\def\eeqa{\end{array}\end{equation}}
\def\eqlab#1{\label{eq:#1}}
\def\figlab#1{\label{fig:#1}}

\def\seclab#1{\label{sec:#1}}
\def\eref#1{(\ref{eq:#1})}
\def\Eqref#1{Eq.~(\ref{eq:#1})}

\def\Figref#1{Fig.~\ref{fig:#1}}

\def\half{\mbox{$\frac{1}{2}$}}

\def\barr{\left(\begin{array}{c}}
\def\earr{\end{array}\right)}
\def\bmat{\left(\begin{array}{cc}}
\def\emat{\end{array}\right)}
\def\al{\alpha}

  \def\eps{\epsilon}

\def\nn{\nonumber}
\def\dd{\mathrm{d}}

\usepackage{tikz}
\usetikzlibrary{decorations.markings}
\usetikzlibrary{decorations.pathmorphing}
\tikzset{photon/.style={decorate, decoration={snake, segment length=2mm, amplitude=1.7pt}}}
\tikzset{boson/.style={decorate, decoration=zigzag}}
\tikzset{fermion/.style={postaction={decorate}, decoration={markings, mark=at position 0.55 with {\arrow{>}}}}}
\usetikzlibrary{calc}
\def\centerarc[#1](#2)(#3:#4:#5){ \draw[#1] ($(#2)+({#5*cos(#3)},{#5*sin(#3)})$) arc (#3:#4:#5); }
\usetikzlibrary{patterns}

\DeclareMathOperator\im{Im}

\def\3d{3-D}

\def\piEFT/{$\slashed{\pi}$EFT}

\def\2PE{2$\upgamma$}

\usepackage{xcolor}
\usepackage{bm}
\usepackage{slashed}
\usepackage{graphicx}
\allowdisplaybreaks

\makeatletter
\g@addto@macro\bfseries{\boldmath}
\makeatother

\definecolor{shadecolor}{cmyk}{0,0,0.45,0}
\definecolor{light-blue}{cmyk}{0.25,0,0,0}
\newsavebox{\mysaveboxM}
\newsavebox{\mysaveboxT}

\begin{document}

 \author{Franziska Hagelstein}
 \author{Vadim Lensky}
 \affiliation{Institut f\"ur Kernphysik and PRISMA$^{++}$ Cluster of Excellence,
 Johannes Gutenberg-Universit\"at  Mainz,  D-55099 Mainz, Germany}

\author{Bogdan Malaescu}
\affiliation{LPNHE, Sorbonne Universit\'e, Universit\'e Paris Cit\'e, CNRS/IN2P3, 75005 Paris, France}

\author{Vladimir Pascalutsa}
 \affiliation{Institut f\"ur Kernphysik,
 Johannes Gutenberg-Universit\"at  Mainz,  D-55099 Mainz, Germany}

\title{Hadronic vacuum polarization in hydrogen-like atoms and ions amid the interplay of recoil and finite-size effects}

\begin{abstract}

Hadronic vacuum polarization (hVP) enters simple
atomic systems at a level that is small yet decisive for the precision
spectroscopy now underway.  We evaluate the hVP contributions to the
Lamb shift and the hyperfine splitting~(HFS) in ordinary and muonic hydrogen (H and $\mu$H) and hydrogen-like helium-3 ions ($^3$He$^+$ and $\mu^3$He$^+$), using the dispersive data-driven approach and state-of-the-art
empirical parametrizations of the $R$ ratio.  At the centre of the analysis is the interplay of
recoil and finite-size effects: the recoil corrections that dominate
the HFS in muonium (Mu), where both constituents are pointlike, are shown
to be suppressed by the nuclear elastic form factors (FFs).  
Our results for the leading hVP contribution to the Lamb shift agree with the
literature within uncertainties. Furthermore, we present a first evaluation of the subleading $O(Z^5\alpha^6)$ hVP--finite-size correction, which is by no means negligible in $\mu^3$He$^+$. Our results for the hVP contribution to the HFS deviate
significantly from all previous evaluations.  For the ground-state HFS, we obtain $2.153(11)$~$\upmu$eV
in $\mu$H and $-15.19(57)$~$\upmu$eV in $\mu^3$He$^+$, as well as
$0.0860(4)$~kHz and $-0.476(17)$~kHz in ordinary H and $^3$He$^+$,
respectively. Notably, our result for $\mu$H differs from previous evaluations
by roughly ten times the experimental precision anticipated by the upcoming
CREMA and FAMU measurements.
\end{abstract}

\date{\today}

\maketitle
\tableofcontents
\section{Introduction}

Recent advances in the spectroscopy of simple atoms have enabled increasingly precise measurements that can serve as stringent tests of the Standard Model and probes of New Physics. Realizing their full potential, however, requires equally precise theoretical predictions, for which the non-perturbative nature of QCD presents a major obstacle. A ubiquitous QCD effect common to virtually all low-energy precision tests is the hadronic vacuum polarization (hVP), which is the primary focus of this work.

The recent update of hVP contributions in muonium (Mu) and  light (muonic) atoms and ions~\cite{Karshenboim:2021jsc} emphasized the importance of a consistent treatment of these effects across different observables, for instance, in the context of a CODATA adjustment of fundamental constants. Here we provide a consistent evaluation for Mu, hydrogen (H), muonic hydrogen ($\mu$H), the helium-3 ion ($^3$He$^+$), and the muonic helium-3 ion ($\mu^3$He$^+$), including the recoil and finite-size effects. 

The timing is not incidental. On the $\mu$H side, the
CREMA~\cite{Amaro:2021goz} and FAMU~\cite{FAMU:2025qkv} collaborations are searching for the nearly-forbidden ground-state
HFS, which will set the experimental precision at the $1$~ppm ($\approx 0.2$~$\upmu$eV) level. The theory predictions have
an uncertainty of at least 30 ppm \cite{Maron:2026wrm} (see also the older \cite{Peset:2016wjq,Antognini:2022xoo,Hagelstein:2023owe}), achieved only by scaling
the H HFS, known to parts in trillion~\cite{Hellwig1970}, to reduce the proton-structure model dependence. 
Our present evaluation of 
the hVP contribution to the $\mu$H HFS
disagrees with existing evaluations by at least 10 ppm (see Table \ref{tab:resultsHFS}), which is significant in view of the aforementioned experiments.  These discrepancies with earlier results are due to two causes: a widely used rescaling of the muonic vacuum polarization ($\mu$VP) contribution that, though adequate for the Lamb shift, misstates the hVP-to-$\mu$VP ratio in the HFS, and calculational errors in one earlier combined hVP--finite-size treatment~\cite{Faustov:1997rc}, which we identify term by term.

On the Mu side, where both constituents are
pointlike and the HFS is a clean QED-plus-hVP observable, the MuSEUM
collaboration at J-PARC~\cite{MuSEUM:2020mzm,MuSEUM:2025cmo} aims to improve the measurement of the ground-state
HFS \cite{Liu:1999iz} towards the ppb level. The QED
prediction for this interval was recently reassessed by
Eides~\cite{Eides:2025veo}, who places its uncertainty at the sub-kHz
level, which is comparable to the hadronic contribution reevaluated here.

Further improvements in the
Lamb shift measurements in light muonic atoms and ions are expected from the
CREMA collaboration, which will test the current state-of-the-art theory 
compilation~\cite{Pachucki:2022tgl} and increase the precision of charge radii extractions. Comprehensive theory compilations of the $2S$ and $2P$ levels---including their fine and hyperfine structure---are available for $\mu$H~\cite{Antognini:2013rsa}, $\mu$D~\cite{Krauth:2015nja}, $\mu^3$He$^+$~\cite{Franke:2017tpc}, and the muonic helium-4 ion~($\mu^4$He$^+$)~\cite{Diepold:2016cxv}.
For the Lamb shift, our results for the hVP contribution confirm the literature while improving its precision. Furthermore, we present a first evaluation of the subleading $O(Z^5\alpha^6)$ hVP--finite-size correction, which is by no means negligible in $\mu^3$He$^+$.

Throughout this study we paid close attention to intricate
cancellations due to recoil and finite-size effects, described for the H HFS by Bodwin and Yennie~\cite{Bodwin:1987mj}.
Their interplay works as follows. In Mu, where both constituents are pointlike, the recoil corrections to the hVP contribution are large: the separate terms carry a logarithm of the mass ratio, $\ln(M/m)$, and both the non-recoil result and the recoil correction are individually much larger than their modest sum. A composite nucleus overturns this picture. Its elastic form factors (FFs) cut the loop integral off at a softer scale and at a much sharper rate than relativistic effects, thereby closing the window from which the logarithm would be built up; the recoil correction is demoted from a logarithmically enhanced effect to an ordinary one, and the full and non-recoil weighting functions almost coincide. There follows a simple criterion, made quantitative in Sec.~\ref{sec:recoil_finite_size}: recoil matters only when the vacuum-polarization (VP) spectral function reaches up to the heavier mass and no FF cutoff intervenes below it.

The paper is organized as follows. Section~\ref{sec:Sec2} sets out the formalism for the dominant hVP, recoil, and finite-size effects.
Section~\ref{sec:Sec3} analyzes the recoil/finite-size interplay, both qualitatively and numerically.  Section~\ref{sec:DHMZ} describes the data-driven evaluation of the hVP based on the Davier, Hoecker, Malaescu, and
Zhang~(DHMZ) parametrization of the $R$ ratio~\cite{Davier:2023fpl}. Section~\ref{sec:results} presents and discusses our results for the Lamb shift and the HFS in $(\mu)$H and $(\mu)^3$He$^+$ in comparison with the literature, and Sec.~\ref{sec:Sec5} concludes. Further numerical details, including a comparison of $R$ ratio inputs, are collected in Appendix~\ref{sec:appendixA}.

\section{Formalism} \seclab{Sec2}
The theoretical description of a hydrogen-like atom in the Standard Model is organized as a systematic expansion around the two-body Coulomb problem. At leading order, the spectrum is that of a lepton bound in the Coulomb field of a pointlike nucleus with charge $Ze$. The Coulomb energies $E^\mathrm{C}$ depend on the principal quantum number $n$ and are degenerate for different values of the orbital (total) angular momentum number $l$ ($j$). The fine and hyperfine structure follow from the Dirac equation and the leading magnetic (spin-spin) interaction. Successive corrections are ordered in powers of the fine-structure constant $\al$ and the lepton-to-nucleus mass ratio $m/M$, and may be grouped as
\beq
E_{nlj}=E^\mathrm{C}+\delta E^\mathrm{QED}+\delta E^\mathrm{recoil}+\delta E^\mathrm{fs}+\delta E^\mathrm{pol}+\ldots\,,
\eqlab{budget}
\eeq
where $\delta E^\mathrm{QED}$ collects the radiative corrections (lepton self-energy and VP), $\delta E^\mathrm{recoil}$ the relativistic corrections associated with the finite nuclear mass, $\delta E^\mathrm{fs}$ the nuclear finite-size effects, and $\delta E^\mathrm{pol}$ the nuclear polarizability entering through two-photon exchange; the ellipsis stands for weak and higher-order terms. Comprehensive accounts of this expansion for electronic and muonic atoms can be found in \cite{Eides:2000xc,Karshenboim:2005iy,Pachucki:2022tgl,Maron:2026wrm}. 

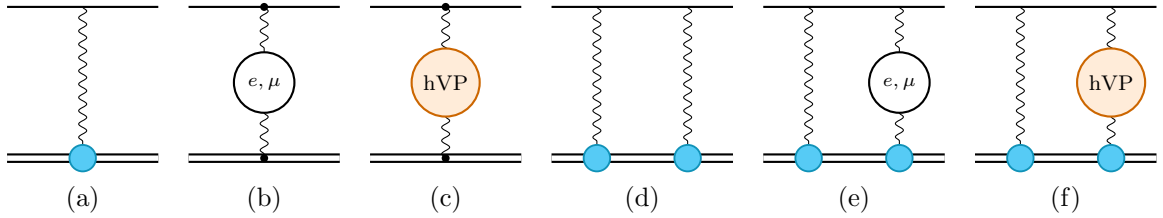
\begin{figure}[t]
\begin{tikzpicture}
\begin{feynman}

  \begin{scope}[shift={(0,0)}]
    \draw[thick] (-1.0,1.0) -- (1.0,1.0);
    \draw[thick, double, double distance=2pt] (-1.0,-1.0) -- (1.0,-1.0);
    \draw[myphoton] (0, 1.0) -- (0,-0.825);
    \node[fsblob] at (0,-1.0) {};
    \node at (0,-1.55) {(a)};
  \end{scope}

  \begin{scope}[shift={(2.4,0)}]
    \draw[thick] (-1.0,1.0) -- (1.0,1.0);
    \draw[thick, double, double distance=2pt] (-1.0,-1.0) -- (1.0,-1.0);
    \draw[myphoton] (0, 1.0)  -- (0, 0.40);
    \draw[thick] (0, 0.0) circle (0.40cm);
    \node[font=\scriptsize] at (0, -0.02) {$e,\mu$};
    \draw[myphoton] (0,-0.40) -- (0,-1.0);
    \fill (0, 1.0) circle (1.5pt);
    \fill (0,-1.0) circle (1.5pt);
    \node at (0,-1.55) {(b)};
  \end{scope}

  \begin{scope}[shift={(4.8,0)}]
    \draw[thick] (-1.0,1.0) -- (1.0,1.0);
    \draw[thick, double, double distance=2pt] (-1.0,-1.0) -- (1.0,-1.0);
    \draw[myphoton] (0, 1.0)  -- (0, 0.35);
    \draw[myphoton] (0,-0.45) -- (0,-1.0);
    \node[hvpblob] at (0, 0.0) {\footnotesize hVP};
    \fill (0, 1.0) circle (1.5pt);
    \fill (0,-1.0) circle (1.5pt);
    \node at (0,-1.55) {(c)};
  \end{scope}

  \begin{scope}[shift={(7.4,0)}]
    \draw[thick] (-1.2,1.0) -- (1.2,1.0);
    \draw[thick, double, double distance=2pt] (-1.2,-1.0) -- (1.2,-1.0);
    \draw[myphoton] (-0.6, 1.0) -- (-0.6,-0.825);
    \draw[myphoton] ( 0.6, 1.0) -- ( 0.6,-0.825);
    \node[fsblob] at (-0.6,-1.0) {};
    \node[fsblob] at ( 0.6,-1.0) {};
    \node at (0,-1.55) {(d)};
  \end{scope}

  \begin{scope}[shift={(10.2,0)}]
    \draw[thick] (-1.2,1.0) -- (1.2,1.0);
    \draw[thick, double, double distance=2pt] (-1.2,-1.0) -- (1.2,-1.0);
    \draw[myphoton] (-0.6, 1.0) -- (-0.6,-0.825);
    \node[fsblob] at (-0.6,-1.0) {};
    \draw[myphoton] (0.6, 1.0)  -- (0.6, 0.40);
    \draw[thick] (0.6, 0.0) circle (0.40cm);
    \node[font=\scriptsize] at (0.6, -0.02) {$e,\mu$};
    \draw[myphoton] (0.6,-0.40) -- (0.6,-0.825);
    \node[fsblob] at (0.6,-1.0) {};
    \node at (0,-1.55) {(e)};
  \end{scope}

  \begin{scope}[shift={(13.0,0)}]
    \draw[thick] (-1.2,1.0) -- (1.2,1.0);
    \draw[thick, double, double distance=2pt] (-1.2,-1.0) -- (1.2,-1.0);
    \draw[myphoton] (-0.6, 1.0) -- (-0.6,-0.825);
    \node[fsblob] at (-0.6,-1.0) {};
    \draw[myphoton] (0.6, 1.0)  -- (0.6, 0.35);
    \draw[myphoton] (0.6,-0.45) -- (0.6,-0.825);
    \node[hvpblob] at (0.6, 0.0) {\footnotesize hVP};
    \node[fsblob] at (0.6,-1.0) {};
    \node at (0,-1.55) {(f)};
  \end{scope}

\end{feynman}
\end{tikzpicture}
\caption{One-photon-exchange (OPE) and two-photon-exchange (TPE) potentials with nuclear finite-size and VP corrections: (a) OPE with electromagnetic form factors, (b) OPE in pointlike nucleus limit with leptonic VP, (c) OPE in pointlike nucleus limit with hVP, (d) TPE with electromagnetic form factors, (e) TPE with leptonic VP, and (f) TPE with hVP. The crossed and time-reversed graphs are not shown.\figlab{VPinterference}}
\end{figure}

To $O(Z^5\alpha^5)$,
the corrections in Eq.~\eqref{eq:budget} can be computed 
via one- and two-photon-exchange diagrams, such as the
ones depicted in Fig.~\ref{fig:VPinterference}. The finite-size effects, encoded in graphs (a) and (d), serve as a template for the VP effects [graphs (b), (c), (e) and (f)] that we treat by close analogy below. The hVP corrections involving the polarizability effects are very small and will not be considered here.

The finite-size effects are parametrized by the elastic electromagnetic FFs of the nucleus. For a spin-1/2 nucleus, these are the Sachs electric and magnetic FFs: $G_E(Q^2)$ and $G_M(Q^2)$, taken as functions of the space-like photon virtuality $Q^2=-q^2\ge 0$ ($q$ the photon momentum), normalized as $G_E(0)=1$ and $G_M(0)=1+\kappa$, with $\kappa$ the anomalous magnetic moment. In hydrogen-like systems, their leading $O(Z^4 \alpha^4)$ effect is characterized
by the root-mean-square (rms) charge radius, $r_E^2=-6\,G_E'(0)$,  contribution to the Lamb shift [from the one-photon exchange graph (a)]. At subleading $O(Z^5 \alpha^5)$, the Zemach and Friar radii [from the two-photon exchange graph (d)] enter the finite-size corrections to the Lamb shift and the HFS, given in the respective subsections below.  Subtractions in the integrals defining these radii through the two-photon exchange, see Eqs.~\eref{ZemachRadius} and \eref{FriarRadius} below, prevent the double-counting of the pointlike and the leading $O(Z^4\alpha^4)$ finite-size contributions and, in practice, render these integrals infrared convergent. Alternatively, the Zemach and Friar radii result from graph (a) treated to second order in Schrödinger perturbation theory.

The VP of the exchanged photon is dominated by its leptonic (QED) part but also receives a hadronic contribution---the hVP. In what follows we evaluate the hVP effect on the Lamb shift and the HFS, treating the nuclear finite-size effects on the same footing and observing their interplay with the recoil corrections.

We begin with the Lorentz- and gauge-invariant  VP tensor
\beq
\Pi^{\mu \nu}(q)=\left(g^{\mu \nu} q^2-q^\mu q^\nu\right)\Pi(q^2),
\eeq
with $\Pi(q^2)$ a scalar function, which 
satisfies 
the once-subtracted dispersion relation:
\beq
\Pi (q^2) = 
\frac{q^2}{\pi} \lim_{\eps\to 0^+} \int\limits^\infty_{s_0} \frac{\dd s}{s}\, \frac{\im  \Pi(s)}{s-q^2 - i\eps}, \label{eq:PiDR}
\eeq 
where $s_0$ is the lowest particle-production threshold (e.g., $s_0=4m_\ell^2$ for a leptonic VP, with $m_\ell$ the lepton mass; or $s_0=4m_\pi^2$, where $m_\pi$ is the charged pion mass, for the hVP). While the leptonic VP contributions are computed perturbatively in QED, 
the hVP contribution is essentially non-perturbative. 
Lattice QCD and the data-driven dispersive approach are 
currently the two best methods to compute it; we use the latter. This means we operate with the $R$ ratio, which represents an empirical cross-section of $e^+e^-$ annihilation into hadrons:
\beq
R(s)=\frac{3s}{4\pi\alpha^2}\sigma(e^+e^- \rightarrow \gamma^* \rightarrow \text{hadrons})\,,
\eeq
where $s$ denotes the squared center-of-mass energy  of the lepton pair. To an excellent approximation, the hVP is then determined by means of an optical theorem,
\beq 
\label{eq:Rfunction}
\im  \Pi^\mathrm{hVP}(s) = -\frac{\al}{3} R(s),
\eeq
substituted in the above dispersion relation.
Note that we are not factoring out $e^2$ or $\al/\pi$ from the VP amplitudes, contrary to other frequently used conventions. 

Below, we provide expressions for the dominant VP contributions to the HFS and Lamb shift, shown in Fig.~\ref{fig:VPinterference} (b), (c), (e) and (f), taking into account the nuclear finite size. Our results 
are written in terms of $\im  \Pi$, but in case of the hVP, we use \eqref{eq:Rfunction} with the empirical parametrizations for the $R$ ratio.

\subsection{Hyperfine Splitting} \seclab{HFSresult}

The leading, non-recoil finite-size effect in the HFS is the Zemach correction,
\beq
E^\mathrm{fs}_{nS\text{-HFS}} = -\,2\,\frac{Z\al\, m_r}{n^3}\, E_\mathrm{F}\, r_\mathrm{Z}\,,\label{eq:FSeffectsHFS}
\eeq
with $m_r=m M/(m+M)$ the system's reduced mass, the Fermi energy given by\footnote{Note that our choice differs from the common definition for Mu, see the discussion in \cite{Karshenboim:2021jsc}.}
\beq
E_\mathrm{F}=\frac{8}{3}\frac{(Z\al)^4 m_r^3}{mM}(1+\kappa)\,,
\eeq
and the Zemach radius:
\beq
r_\mathrm{Z}=-\frac{4}{\pi}\int\limits_0^\infty\frac{\dd Q}{Q^2}\left(\frac{G_E(Q^2)\,G_M(Q^2)}{1+\kappa}-1\right).\eqlab{ZemachRadius}
\eeq

The master formula for the $O(Z^5 \alpha^6)$ VP contribution to the HFS, overlayed with recoil and finite-size effects, takes the following form:
\bea
E_{nS\text{-HFS}}^{\text{fs-VP}}&=&-E_\mathrm{F} \frac{2Z\alpha}{\pi^2n^3}\frac{mM}{M^2-m^2}\int\limits_{t_0}^\infty 
\frac{\dd t}{t}\,W(t)\, \im \Pi(t)\label{eq:finalMasterformula},
\eea
with the weighting function given by (see, e.g.,~\cite[Sec.~6]{Hagelstein:2015egb})
\beq
W(t)=\frac{1}{1+\kappa}\int\limits_0^\infty \frac{\dd Q}{Q}\bigg\{2(v-v_l) G_M(Q^2)\bigg[2F_1(Q^2)+\frac{F_1(Q^2)+3F_2(Q^2)}{(v_l+1)(v+1)}  \bigg]-\left[1-\frac{m^2}{M^2}\right]\frac{5+4v_l}{(1+v_l)^2}F_2^2(Q^2)\bigg\}
\frac{Q^2}{t+Q^2}\,,
\label{eq:elasticHFS}
\eeq
with the dimensionless quantities $\tau=Q^2/4M^2$, $\tau_l=Q^2/4m^2$,
$v=\sqrt{1+\tau^{-1}}$, and $v_l=\sqrt{1+\tau_l^{-1}}$. The Dirac and Pauli FFs are expressed in terms of the Sachs FFs by 
\beq F_1(Q^2)=\frac{1}{1+\tau}\left[G_E(Q^2)+\tau G_M(Q^2)\right], \qquad F_2(Q^2)=\frac{1}{1+\tau}\left[G_M(Q^2)-G_E(Q^2)\right].
\eeq

Expanding $W(t)$ in powers of $1/M$ and $1/m$ for large masses, the leading term recovers the non-recoil limit: 
\beq
W_\text{non-recoil}(t)=\frac{8(M-m)}{1+\kappa}\int\limits_0^\infty \dd Q
\frac{G_M(Q^2)G_E(Q^2)}{t+Q^2}\,.
\label{eq:elasticHFSnorecoil}
\eeq
 Its regime of applicability is limited formally to cases where both masses ($M$ and $m$) are  much larger than the typical values of $Q$ intrinsic to the nuclear FFs.

In the limit of pointlike FFs, where $F_1\rightarrow 1$ and $F_2\rightarrow \kappa$, the weighting function takes the form
\begin{equation}
W^\odot(t)=\left[(3\kappa+1)b+2\right]\sqrt{1-1/b}\, \ln \frac{1-\sqrt{1-1/b}}{1+\sqrt{1-1/b}}+\left[(3\kappa+1)b-\frac{3}{2}(\kappa
   -1)\right] \ln 4 b-\frac{3\kappa+1}{2}-(M\to m),\,
\label{eq:pointlike}
\end{equation}
with $b=t/4M^2$. In this limit, the last term in Eq.\ \eqref{eq:elasticHFS}, proportional to $F^2_2(Q^2)$, would superficially give a divergent integral; however, it actually cancels with an analogous term generated by the polarisability contribution (see discussions in \cite{Hagelstein:2023owe,Ruth:2024bsl}) once the VP effects are taken into account there.

Furthermore, when the anomalous magnetic moment is small (which is the case for leptons but not nuclei), $\kappa\to 0$, the weighting function simplifies to the result of \citet{Sapirstein:1983xr}: 
\beq
 W^{\odot\slashed{\kappa}}(t)=\left(b+2\right)\sqrt{1-1/b}\,\ln\frac{1-\sqrt{1-1/b}}{1+\sqrt{1-1/b}}+\left(b+\nicefrac{3}{2}\right)\ln 4b-\nicefrac{1}{2}-(M\to m)\,,\label{eq:kernelpoint}
\eeq
to which we refer here as the structureless limit.
The expansion of $W^{\odot}(t)$ in powers of $1/M$ and $1/m$ gives
\beq
\label{eq:expandedHFS}
W^{\odot}(t)=\frac{4\pi}{\sqrt{t}}(M-m)+\frac{3\pi}{2}\kappa\sqrt{t}\left(\frac{1}{M}-\frac{1}{m}\right)+3(\kappa-1)\ln \frac{M}{m}+O(t/M^2,t/m^2)\,,
\eeq
where the first term corresponds to the combined non-recoil pointlike limit:
\beq
\label{eq:pointlikenorecoilHFS}
W^{\odot}_\text{non-recoil}(t)=\frac{4\pi}{\sqrt{t}}(M-m).
\eeq

\subsection{Lamb Shift}\seclab{LSresult}
The leading, non-recoil finite-size effect in the Lamb shift is very well-known:
\beq
E^\mathrm{fs}_{nS} = \frac{2}{3}\,\frac{(Z\al)^4 m_r^3}{n^3}\left[r_E^2 - \half\, Z\al\, m_r\, r_\mathrm{F}^3\right],\label{eq:FSeffectsLS}
\eeq
with the Friar radius
\beq
r_\mathrm{F}^3=\frac{48}{\pi}\int\limits_0^\infty\frac{\dd Q}{Q^4}\left[G_E^2(Q^2)-1+\tfrac{1}{3}r_E^2 Q^2\right].\eqlab{FriarRadius}
\eeq

As noted above, contrary to the HFS, the VP corrections to the Lamb shift start already at $O(Z^4\al^5)$ through the Uehling-type potential generated by one insertion of the VP in the photon propagator (see, e.g.,~\cite{Antognini:2022xoo}):
\begin{subequations}
\label{eq:ELSmr}
\bea
E_{nS}^{\text{VP}} &=&\begin{dcases} \frac{4(Z\al)^4 m_r^3}{\pi} \int_{t_0}^\infty  \frac{\dd t}{t} \, \frac{\im \Pi (t)}{(\sqrt{t}+ 2Z \al m_r)^2}\,,&n=1,\\
  \frac{(Z\al)^4 m_r^3}{4\pi} \int_{t_0}^\infty  \frac{\dd t}{t} \, \frac{\left[(Z\alpha m_r)^2+2t\right]\im \Pi (t)}{(\sqrt{t}+ Z \al m_r)^4}\,, &n=2
  \end{dcases}\label{eq:LOLSexact}\\
  &=& 
\frac{4(Z\al)^4 m_r^3}{n^3}\, \Pi'(0) + O( Z^5\al^6). \label{eq:LOLSapprox}
\eea
\end{subequations}
This is similar to the leading hVP contribution to the electron anomalous magnetic moment $a_e=\half (g_e-2)$:
\beq\label{eq:ae}
a_e^\text{hVP}\simeq -\frac{\alpha}{3\pi} m_e^2\, \Pi^{\prime}(0).
\eeq
Here, $\Pi'(0)$ is the slope of the scalar VP function at $q^2=0$. Following \Eqref{PiDR}, it can be written as
\beq 
\Pi'(0) =\frac{\dd \Pi(q^2)}{\dd q^2}\bigg\vert_{q^2=0}= \frac{1}{\pi}\int_{t_0}^\infty \!\dd t\, \frac{ \im\Pi(t)}{t^{2}}.
\eeq 
The expansion of \Eqref{ELSmr} parallels the expansion of finite-size effects in Eq.\ \eqref{eq:FSeffectsLS} that gives the leading contribution proportional to the charge radius squared. 
This equation represents a pure VP effect,\footnote{It is important to note that the expansion of \Eqref{ELSmr} in powers of $Z\alpha m_r$, see \Eqref{LOLSapprox}, is not applicable to eVP contributions in muonic atoms. Because $Z\al m_r \centernot\ll \sqrt{t_0} =2m_e$, the eVP contributions are enhanced in these systems. Consequently, \Eqref{ELSmr} with a one-loop eVP insertion cannot be regarded as a $O(Z^4 \alpha^5)$ correction, but is formally counted as $O(Z^2 \alpha^3)$ \cite{Pachucki:2022tgl}.} with the leading $O(Z^4\alpha^5)$ term proportional to the slope of the scalar VP function, and the $O(Z^5\alpha^6)$ correction to it being very small, as shown in Table \ref{tab:expcoeff}. 
The mixed VP and finite-size corrections to the $nS$ Lamb shift are also subleading $O(Z^5\alpha^6)$. They are defined analogously to the HFS:
\bea
E_{nS}^{\text{fs-VP}}&=&-16\,\frac{m}{M}\frac{(Z\al)^5 m_r^3}{\pi^2 n^3}\int\limits_{t_0}^\infty \frac{\dd t}{t^2}\,U(t)\, \im \Pi(t),\label{eq:finalMasterformulaLS}
\eea
in terms of the weighting functions
\begin{subequations}
\bea
U(t)&=&\int_0^\infty \frac{\dd Q}{Q}\,\Bigg(\frac{1}{(1+\tau)(v_l+v)}\Bigg\{\left[\tau+\frac{3+2\tau}{(1+v_l)(1+v)}\right]G_M^2(Q^2)\nn\\
&-&\frac{1}{\tau}\left[1-\frac{1}{(1+v_l)(1+v)}\right]\left[G_E^2(Q^2)-1\right] \Bigg\}-\frac{v_l+2}{(1+v_l)^2}\,\left[F_1^2(Q^2)-1\right]\Bigg)\frac{t}{t+Q^2},\quad\label{eq:LSweighting}\\
U_\text{non-recoil}(t)&=&-\frac{m_r}{m}\int_0^\infty \frac{\dd Q}{Q}\frac{2M\left[G_E^2(Q^2)-1\right]}{Q}\frac{t}{t+Q^2},\label{eq:LSweightingnonrecoil}\\
U^{\odot}(t)&=&(1+\kappa)^2\int_0^\infty \frac{\dd Q}{Q}\,\frac{1}{(1+\tau)(v_l+v)}\left[\tau+\frac{3+2\tau}{(1+v_l)(1+v)}\right]\frac{t}{t+Q^2},\quad\label{eq:LSweightingpointlike}
\eea
\end{subequations}
where $U(t)$ is the full kernel,  $U_\text{non-recoil}(t)$ corresponds to the non-recoil limit, and $U^{\odot}(t)$ corresponds to the pointlike limit. The combined non-recoil pointlike limit is covered by \Eqref{ELSmr}.

\section{Recoil versus Finite-Size Effects in the Hyperfine Splitting} \seclab{Sec3}
\label{sec:recoil_finite_size}
Before considering the results for $(\mu)$H and $(\mu)^3$He$^+$, let us discuss the interplay of different scales which govern the recoil and finite-size corrections. We start by comparing the HFS in Mu and ($\mu$)H. In the former, recoil corrections to the hVP contributions are important, whereas in the latter, the FFs render these corrections nearly negligible.
\begin{figure}[t]
\begin{tabular}{cc}
\includegraphics[height=0.35\textwidth]{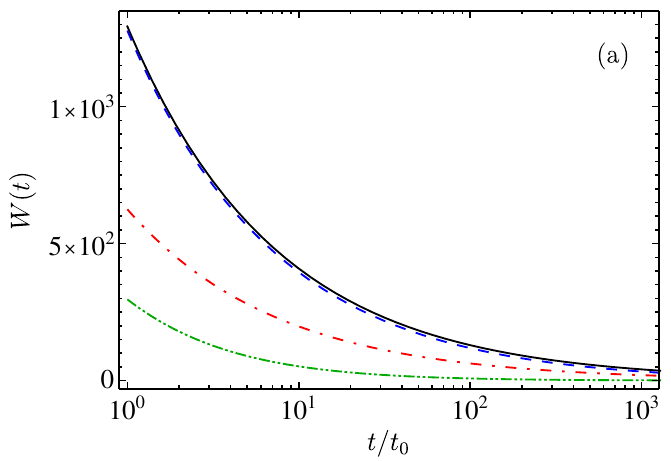} &
\includegraphics[height=0.35\textwidth]{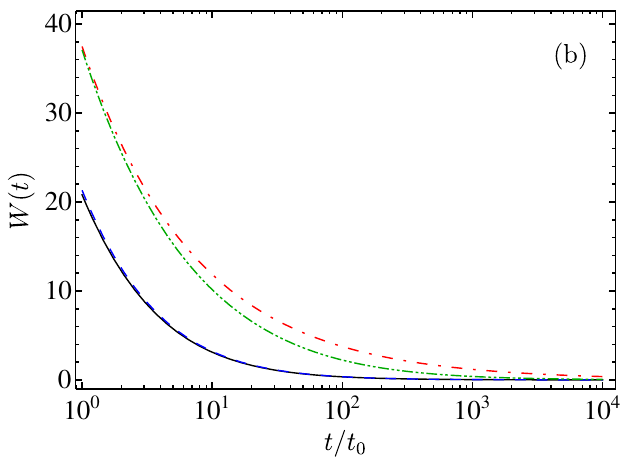}
\end{tabular}
\caption{Weighting functions in Mu and $\mu$H, showing the effect of recoil, its dependence on the scales of the polarization operator, and the finite-size effects. Panel (a): $W(t)$ in Mu, shown as a function of the rescaled variable $t/t_0$. Black solid and blue dashed: $W^\odot_\mathrm{non-recoil}(t)$ and $W^\odot(t)$ for $t_0=4m_e^2$. Red dot-dashed and green dot-dot-dashed: $W^\odot_\mathrm{non-recoil}(t)$ and $W^\odot(t)$ for $t_0=4m_\mu^2$. Panel (b): $W(t)$ in $\mu$H, shown as a function of $t/t_0$ for $t_0=4m_\pi^2$. Black solid and blue dashed: $W(t)$ and $W_\mathrm{non-recoil}(t)$. Red dot-dashed and green dot-dot-dashed: $W^\odot_\mathrm{non-recoil}(t)$ and $W^\odot(t)$. 
\label{fig:wfs}
}
\end{figure}

To see this, note that $W^\odot_\mathrm{non-recoil}(t)$ is just the first term in the expansion of $W^\odot(t)$ in powers of $t$ (at small $t$). The relative difference between the two is small at $t\to 0$ and thus the recoil corrections are suppressed in this regime. On the other hand, $W^\odot_\mathrm{non-recoil}(t)$ and $W^\odot(t)$ exhibit distinct asymptotic behaviors at $t\to\infty$. Namely, Eq.\ \eqref{eq:pointlikenorecoilHFS} implies that $W^\odot_\mathrm{non-recoil}(t)=O(t^{-1/2})$ in this limit, whereas $W^\odot(t)=O(t^{-1}\ln t)$, as follows from Eq.\ \eqref{eq:pointlike}. In this regime (or, more generally, outside of the non-recoil regime) the recoil corrections to the weighting function are formally of the same size as the function itself and therefore need to be included. We also note that in a system where $m\ll M$, such as Mu, the contribution of the $(M\to m)$ term in Eq.\ \eqref{eq:pointlike} is negligibly small (of the relative order of $\sim 10^{-4}$ for Mu). 

Identifying which one of the two regimes above applies to a particular VP can be done by comparing the characteristic VP scale (typically the threshold value $\sqrt{t_0}$) with the mass of the heavier particle $M$; a bigger ratio leads to bigger recoil effects. To illustrate this, panel (a) of Fig.\ \ref{fig:wfs} shows $W^\odot(t)$ and $W^\odot_\mathrm{non-recoil}(t)$ for Mu as functions of the rescaled variable $t/t_0$ for eVP and $\mu$VP (corresponding to $t_0=4m_e^2$ and $t_0=4m_\mu^2$). One can see that eVP clearly corresponds to a very small recoil correction, whereas in the case of $\mu$VP the recoil correction is of the same order of magnitude as the non-recoil weighting function. This is in full accordance with the considerations above: $\sqrt{t_0}=2m_e\ll m_\mu$ corresponds to the non-recoil regime for eVP, whereas $\mu$VP calls for using the full $W^\odot(t)$. The latter naturally applies also to hVP, with its typical scales being even larger than $m_\mu$.

It is instructive to trace where the recoil correction acquires its size. Expanding the pointlike kernel, at large masses, Eq.\ \eqref{eq:expandedHFS}, the leading term reproduces the non-recoil result. The first genuine recoil correction is the $t$-independent piece $3(\kappa-1)\ln M/m$, which carries the logarithm of the lepton-to-nucleus mass ratio. It is this logarithmic enhancement that renders the recoil correction large whenever the VP spectral function reaches up to the heavier mass. The mechanism is the same one identified long ago in the recoil corrections to the H HFS by Bodwin and Yennie~\cite{Bodwin:1987mj}: there the individual contributions are each enhanced by $\ln(M/m)$ and are individually much larger than their sum, so that the modest net result, including recoil corrections, survives only as the residue of a delicate cancellation. In structureless Mu the enhancement is striking---the recoil correction pulls the weighting-function integral down from its non-recoil value of $1.237(6)$~kHz to $0.2333(11)$~kHz (Table~\ref{tab:expcoeff2}), a suppression by more than a factor of five, with $\ln(m_\mu/m_e)$ being the main cause.

This situation is completely altered by the finite-size effects. The small $t$ regime is again the same for $W(t)$ and $W_\mathrm{non-recoil}(t)$: $W(t)\simeq W_\mathrm{non-recoil}(t)=O(t^{-1/2})$ [which can be traced from Eq.\ \eqref{eq:elasticHFS} or, respectively, Eq.\ \eqref{eq:elasticHFSnorecoil} using integration by parts]. However, as is straightforward to see, both $W(t)$ and $W_\mathrm{non-recoil}(t)$ are $O(t^{-1})$ at large $t$. The importance of the recoil terms is thus greatly reduced in general by the inclusion of the finite-size effects, which is to be expected since the elastic FFs suppress the contribution of high $Q$.

Put differently, recoil and finite size act as two competing cutoffs on the high-$Q$ region: relativistic recoil at the nuclear mass scale $M$, and the elastic FFs at the lower scale $\Lambda\sim m_\rho$ (lower still for a composite nucleus, such as the helion, whose coherent FF cuts off at the inverse nuclear radius). Since $\Lambda\lesssim M$, the softer cutoff prevails, and the recoil logarithm never develops. It is worth stressing that ``softer'' here refers to the {\it scale} of the cutoff, not to the steepness of the fall-off: the $\sim 1/Q^4$ decay of the FFs is in fact faster in $Q$ than the merely logarithmic sensitivity of the recoil kernel.

This reasoning furnishes a simple criterion for when the recoil corrections to a VP contribution may be neglected: they are logarithmically enhanced only if the VP spectral function extends to scales comparable with, or above, the mass of the heavier constituent, and only so long as no FF cutoff intervenes at a lower scale. For the hVP in Mu both conditions hold and recoil dominates; for the hVP in $\mu$H, H, and $(\mu)^3$He$^+$ the nuclear FFs act at $Q\sim\Lambda\ll M$, the first condition is thereby rendered moot, and the recoil correction is negligible. The same criterion accounts for the contrast, visible in panel~(a) of Fig.~\ref{fig:wfs}, between the tiny eVP recoil correction in Mu---where $\sqrt{t_0}=2m_e\ll m_\mu$ keeps the integral in the non-recoil regime---and the sizeable $\mu$VP one.

To demonstrate that, panel (b) of Fig.\ \ref{fig:wfs} shows the weighting functions for $\mu$H, both in the pointlike limit and including the finite-size effects, as functions of the rescaled variable $t/t_0$ for the case of hVP, i.e., $t_0=4m_\pi^2$. The finite-size effects are calculated using the proton elastic FFs extracted in \cite{Borah:2020gte}. One can see that $W(t)$ and $W_\mathrm{non-recoil}(t)$ are on top of each other, whereas the difference between $W^\odot(t)$ and $W^\odot_\mathrm{non-recoil}(t)$ is sizeable, especially at higher values of $t/t_0$. In fact, one can see that at very small $t\simeq t_0$ the non-recoil regime works well also in the pointlike case, however, the recoil correction is still rather sizeable [recall that $R(t_0)=0$ and $R(t)$ is very large in the $\rho$ region ($t/t_0\simeq 10$), and also has a rather extended high-$t$ tail]. 

Figure \ref{fig:wfs} also shows how the finite-size effects generally suppress the total hVP contribution to the HFS in $\mu$H and the high-$t$ tail of $W(t)$. The smallness of the recoil correction, as well as the suppression of the total hVP contribution by the elastic FFs, is completely analogous to the behaviour seen in H and $(\mu)^3$He$^+$, with the details defined by the different mass ratios, magnetic moments, and FFs. To provide further details of the interplay of the recoil corrections and the finite-size effects, Table \ref{tab:expcoeff2} in the Appendix shows the results for the HFS contributions in Mu, $(\mu)$H and $(\mu)^3$He$^+$, corresponding to the different limits for $W(t)$ considered above.

\begin{figure}[t]
\begin{tabular}{cc}
\includegraphics[height=0.35\textwidth]{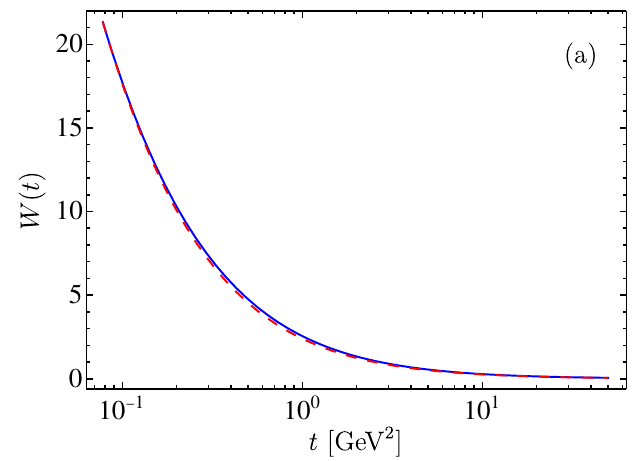} &
\includegraphics[height=0.35\textwidth]{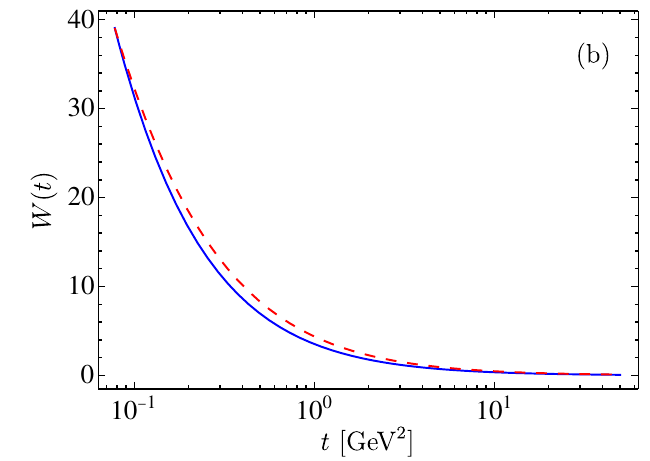}
\end{tabular}
\caption{Comparison of weighting functions for the VP contribution to the HFS [$W(t)$ in \Eqref{elasticHFS}], shown by the solid  blue curves, and the VP contribution to the muon anomalous magnetic moment [$K_a(t)$ in \Eqref{Kweighting}, rescaled to coincide with the respective $W(t)$ at $t=4m_\pi^2$], shown by the red dashed curves. Panels (a) $\mu$H, and (b)  $\mu^3$He$^+$.
\label{fig:wfs_kamu}
}
\end{figure}

To conclude this section, Fig.\ \ref{fig:wfs_kamu} shows the comparison of $W(t)$ for $\mu$H and $\mu^3$He$^+$ with the weighting function for the muon anomalous magnetic moment $a_\mu=\half (g_\mu-2)$,
\begin{equation}
\eqlab{Kweighting}
K_a(t)=\frac{8b^2-8b+1}{\sqrt{1-1/b}}\ln\frac{1-\sqrt{1-1/b}}{1+\sqrt{1-1/b}}+4b\left(2b-1\right)\ln 4b-4b+\frac{1}{2},
\end{equation}
with $b=t/4m_\mu^2$ (see, e.g., \cite{Li:2026} for the context). 
In this figure, we rescale $K_a(t)$ so that it coincides with the respective $W(t)$ at $t=4m_\pi^2$ (the respective factors are $704.71$ for $\mu$H and $1932.3$ for $\mu{^3}$He$^+$). As in Fig.~\ref{fig:wfs}, the proton elastic FFs entering $W(t)$ are from \cite{Borah:2020gte}, whereas those for ${}^3$He are taken from \cite{Amroun:1994qj}.
One can see that $W(t)$ follows the rescaled $K_a(t)$ quite closely, especially for $\mu$H, starting from the threshold and going through the whole region of $t$ important for the HFS contribution (up to a few GeV$^2$). A qualitatively very similar behaviour holds also for the HFS in Mu and both H and ${^3}$He$^+$, not shown here.

\section{Data-driven Evaluation of Hadronic Vacuum Polarization}
\label{sec:DHMZ}

Our analysis relies on the DHMZ methodology to combine experimental cross-section measurements of $e^+e^-$ annihilation into hadrons~(see Refs.~\cite{Davier:2010rnx,Davier:2010nc,davier:2019can} and references therein), originally developed in the context of the Standard Model prediction of the hVP contribution to the muon anomalous magnetic moment, $a_\mu^\mathrm{hVP}$. As is widely recognized, the data-driven dispersive evaluation of the hVP currently faces  tensions among different $e^+e^-$ experimental datasets, particularly in the $\pi^+\pi^-$ channel. 
This has triggered various theoretical efforts, such as the comparison and advancement of state-of-the-art radiative corrections for $e^+e^-$ scattering~\cite{Aliberti:2024fpq}, as well as experimental and phenomenological efforts to study the reliability of these corrections and their actual impact on various experimental analyses~\cite{BaBar:2023xiy,Davier:2023fpl,Belle-II:2024msd}. 
At the same time, recent lattice QCD 
predictions of the hVP contribution are in notable discrepancy with the 
data-driven evaluation. 
These internal and external tensions have motivated 
the use of a combination of lattice QCD results for $a_\mu^\mathrm{hVP}$ in 
the 2025 White Paper of the ``Muon $g-2$ Theory Initiative'' \cite{Aliberti:2025beg}, 
updating the data-driven consensus from \cite{Aoyama:2020ynm}. 
As an outcome of this update, the Standard Model prediction of $a_\mu$ is now consistent with 
the latest experimental measurements from Fermilab \cite{Muong-2:2025xyk}.

The weighting functions of the hVP contributions to the HFS in hydrogen-like atoms and the muon anomalous magnetic moment, displayed in \Figref{wfs_kamu}, are similar in shape and dominated by the lightest hadronic channels. While this opens interesting perspectives for joint studies of hVP effects across the different low-energy observables, it also means that the experimental tensions described above affect our 
results in an analogous manner. It is therefore crucial not only to propagate 
the standard experimental uncertainties of the cross-section data, but also to 
rigorously evaluate the impact of these global data inconsistencies on our 
observables. The DHMZ framework is uniquely equipped to address these exact 
challenges.

The DHMZ approach enables combining cross-section data with arbitrary point spacing or binning, employing spline-based interpolations to redistribute them in a fine common binning. The average weights, derived through a $\chi^2$ minimisation, also account for the different bin sizes and point-spacings of measurements. In addition, the $\chi^2$ evaluation enables a local test of the level of agreement among the input measurements. The averaging procedure has been validated through a closure test~\cite{Davier:2010rnx}. This procedure also provides a full treatment of uncertainties and correlations, between the measurements~(data points or bins) of a given experiment, between experiments and between different channels, while also accounting for systematic tensions among experiments~\cite{Davier:2010rnx,Davier:2010nc,davier:2019can}. 
We use its most recent update~\cite{Davier:2023fpl}, employing in particular the full set of measurements currently available in the $\pi^+\pi^-$ channel, complemented with methods introduced in  \cite{Boccaletti:2024guq} to account for the spread of the input cross-sections. In the $\pi^+\pi^-$ channel, the average is dominated by the most precise experiments (BaBar~\cite{BaBar:2009wpw,BaBar:2012bdw}, CMD3~\cite{CMD-3:2023rfe,CMD-3:2023alj}, KLOE~\cite{KLOE:2008fmq,KLOE:2010qei,KLOE:2012anl}, SND20~\cite{SND:2020nwa}, followed by CMD-2~\cite{Akhmetshin:2001ig,Akhmetshin:2006bx,Akhmetshin:2006wh,Aulchenko:2006na}, BESIII~\cite{BESIII:2015equ} and SND06~\cite{Achasov:2005rg}), while BaBar covers the full energy range of interest. However, some tensions exist in this channel, especially between KLOE and CMD3, which provide the smallest and respectively largest cross-sections in the $\rho$ peak region. They are quantified, e.g., through a local $\chi^2/\rm{ndof}$ in the combination or through the significance of the pairwise difference of $a_\mu^\mathrm{hVP}$ integrals computed for a scan of mass intervals~\cite{Davier:2023fpl}. 
In the current data combination, we complement the local uncertainty rescaling on the basis of the $\chi^2/{\rm ndof}$~\cite{Davier:2010rnx,Davier:2010nc,davier:2019can,Davier:2023fpl} with several uncertainties added to account for the systematic tensions among experiments (i.e., their ``systematic scatter'').
These correspond to the ``BaBar-KLOE'' systematic uncertainty associated to the tension between these measurements~\cite{davier:2019can}, and the uncertainties introduced in  \cite{Boccaletti:2024guq} to conservatively account for the effect of CMD3 on the average, or even the effect of experiments with reduced coverage of the energy range. 
In  \cite{Boccaletti:2024guq}, these uncertainties were accounted for in a hybrid approach based on lattice QCD and dispersive integrals, yielding a theoretical prediction in excellent agreement with the experimental measurement of $a_\mu$ \cite{Muong-2:2025xyk}, while being also in good agreement with and more precise than the combination of lattice QCD results for $a_\mu^\mathrm{hVP}$ from  \cite{Aliberti:2025beg}. However, taking into account the systematic scatter in a stand-alone data-driven analysis is not enough to eliminate the discrepancy between the experimental and theoretical values of $a_\mu$ (which is presumed to be due to the hVP contribution). 
To overcome that obstacle, one would have to inflate the uncertainty of $a^\mathrm{hVP}_\mu$ by an additional factor of $2.44$.

\begin{table*}[tb]
    \centering
    \begin{threeparttable}
        \caption{The hVP contribution to the ground-state hyperfine splitting in Mu. Results previously reported in the literature are shown in the top block. Our results correspond to Eq.\ \eqref{eq:kernelpoint}, i.e., include the recoil corrections. The uncertainty of our result is solely due to the uncertainty of the employed parametrization of $R(t)$.            \label{tab:resultsHFSmuonium}
    }
    \begin{tabular}{l||S[table-format=2.10,group-minimum-digits=1]}
 &\multicolumn{1}{c}{Mu [kHz]}\\
    \hline
       \hline
       Previous calculations:&\\
\quad\citet{Sapirstein:1983xr}   &0.22(3)\\  
\quad\citet{Faustov:1999nty}     &0.2397(70)\\  
\quad\citet{Czarnecki:2001yx}    &0.233(3)\\  
\quad\citet{Nomura:2012sb}       &0.23268(144)\\ 
\quad\citet{Keshavarzi:2019abf}  &0.23204(82)\\ 
\quad\citet{Karshenboim:2021jsc} &0.236(5)\\
        \hline
  {This work} &0.2333(11)    \\
        \hline
    \end{tabular}
    \end{threeparttable}
\end{table*}

To facilitate a comparison of the DHMZ evaluation with older parametrizations of $R(t)$, we start from the hVP contribution to the ground-state
HFS in Mu. Table \ref{tab:resultsHFSmuonium} demonstrates that our results coincide 
with other existing evaluations within uncertainties. 
However, our quoted uncertainty stems exclusively from the standard experimental errors encoded in the parametrization of $R(t)$~(which nevertheless includes the local uncertainty enhancement on the basis of the $\chi^2/{\rm ndof}$)
and does not yet account for the systematic
tensions discussed above. 
To estimate the full uncertainty of our data-driven dispersive evaluation, we follow the same procedure as introduced in \cite{Davier:2023fpl,Boccaletti:2024guq} for the evaluation of  $a_\mu^\mathrm{hVP}$. 
As explained above, a way to reconcile the different experiments is to interpret the systematic scatter between them as another systematic uncertainty. 
Applying the same 
procedure here first increases the quoted error for Mu by a factor of $2.54$ 
(obtained by adding in quadrature the $R(t)$ uncertainty and the effect of the systematic scatter). 
This intermediate error is then scaled by the aforementioned factor of $2.44$ to account 
for the $a_\mu$ discrepancy.
 Using the same factor of $2.44$ for all the contributions considered here is justified by the shape of their $W(t)$ being very similar to $K_a(t)$, as shown in Fig.~\ref{fig:wfs_kamu}. Note that \cite{Karshenboim:2021jsc} used a similar procedure to evaluate their uncertainty shown in Table \ref{tab:resultsHFSmuonium}, neglecting the scatter between different $e^+e^-$ experiments but taking into account the discrepancy between the (average) theoretical and experimentally measured values of $a_\mu$.

\section{Results and Discussion}
\label{sec:results}

In this section, we present our results for the combined hVP and finite-size corrections to the spectra of $(\mu)$H and $(\mu)^3$He$^+$. The results for the ground-state 
HFS and Lamb shift are summarized in Tables \ref{tab:resultsHFS} and 
\ref{tab:expcoeff}, respectively. Further details regarding the HFS and 
Lamb shift contributions, evaluated in the different limits discussed in Sec.\ 
\ref{sec:recoil_finite_size}, are provided in Tables \ref{tab:expcoeff2} 
and \ref{tab:expcoeffall} of Appendix \ref{sec:appendixA}. These tables 
also include a comparison with an evaluation based on the \textsc{alphaQED23} 
parametrization of $R(t)$ \cite{Jegerlehner:2017zsb,Jegerlehner:2019lxt}.

The finite-size effects in H and $\mu$H are calculated using the proton elastic FFs determined in \cite{Borah:2020gte} from a $z$-expansion fit to electron-proton scattering, which imposes the precise proton charge radius  extracted from the $\mu$H Lamb shift \cite{Pohl:2010zza,Antognini:1900ns}. The fit parameters' covariances are employed to obtain the related error estimates, which are added in quadrature with the uncertainty coming from the $R$ ratio, the latter being the dominant one.

The proton finite-size effects, such as the $O(Z^5 \al^5)$ Zemach radius and recoil corrections, bring in an additional large model dependence due to the proton FF parametrization. For example, the present state-of-the-art \cite{Antognini:2022xqf}, based on a dispersion-theoretical analyses of the proton FFs \cite{Lin:2021xrc,Lin:2021umz}, leads to a $\sim 3.7\,\%$ larger energy correction of the HFS in ($\mu$)H than the parametrization of Borah {\it et al.}~\cite{Borah:2020gte,Pachucki:2022tgl} that we use here. On this basis, we assign a $4 \%$ model uncertainty to our calculations in $(\mu)$H involving the FFs.
This uncertainty is not included in the tables below.
It is customary to reduce the hadronic uncertainties in the prediction of the $\mu$H HFS by employing the experimental value of the ground-state HFS in H, known to ppt precision \cite{Hellwig1970}, and relying only on a theoretical prediction of the scaled difference of hadronic corrections in H and $\mu$H \cite{Peset:2016wjq,Antognini:2022xoo,Maron:2026wrm}.

For ${^3}$He$^+$ ions, we use the empirical charge and magnetic FFs of the helion given in \cite{Amroun:1994qj}. As an alternative, we use the FFs obtained in \cite{Piarulli:2012bn} in a chiral effective theory framework and parametrize them using a sum of Gaussians analogously to \cite{Amroun:1994qj}. Here, we take the average of and the scatter between the results obtained with the two FF parametrizations as our result and error estimate. The scatter is then added in quadrature to the uncertainty coming from the DHMZ parametrization of $R(t)$. In both ${^3}$He$^+$ and $\mu{^3}$He$^+$, the latter uncertainty is negligibly small in comparison with the scatter between the two FF parametrizations.

\begin{table*}[t]
    \centering
    \begin{threeparttable}
        \caption{The hVP contribution to the ground-state hyperfine splitting in H, $^3$He$^+$, $\mu$H, and $\mu^{3}$He$^+$. Results previously reported in the literature are shown in the top block. 
        The uncertainty of the pointlike result is solely due to the uncertainty of the employed parametrization of $R(t)$. For the finite-size result, the uncertainties stemming from $R(t)$ and from the elastic FFs are added in quadrature (see text). 
            \label{tab:resultsHFS}
    }
    \begin{tabular}{l||S[table-format=2.8,group-minimum-digits=1]|S[table-format=-2.7,group-minimum-digits=1]|S[table-format=2.7,group-minimum-digits=1]|S[table-format=-3.4,group-minimum-digits=1]}
 &\multicolumn{1}{c|}{H [kHz]}&\multicolumn{1}{c|}{$^3$He$^+$ [kHz]}&\multicolumn{1}{c|}{$\mu$H [$\upmu$eV]} &  \multicolumn{1}{c}{$\mu^3$He$^+$ [$\upmu$eV]}\\
    \hline
       \hline
       Previous calculations:&&&&\\
\quad\citet{Karshenboim:1996ew} &&&&\\
\qquad finite size &0.04(1)&&&\\
\qquad pointlike  &0.19(8)&&&\\
\quad\citet{Faustov:1997rc}&&&3.5610\hspace{-3.ex}
&\\
\quad\citet{Borie:2011eia}&&&4.8(8)&-72.8\\
        \hline
  {This work:}    &&&&\\
  \qquad finite size \eref{elasticHFS} &0.0860(4)&-0.476(17)&2.153(11)&-15.19(57)  \\
  \qquad pointlike  \eref{pointlike} 
      &0.2736(12)&-1.860(10)&7.04(3)&-61.0(3)\\
        \hline
    \end{tabular}
    \end{threeparttable}
\end{table*}

In Table \ref{tab:resultsHFS}, we show our results for the hVP contribution to the ground-state HFS in $(\mu)$H and $(\mu)^3$He$^+$, compared with the results in the literature.
As specified above, our uncertainties correspond to the DHMZ parametrization of $R(t)$ and the nuclear FFs, added in quadrature (except for the model dependence of the proton FFs). Accounting for the experimental scatter in the $e^+e^-$ data would give an extra factor of $2.51$ for $(\mu)$H and $1.06$ for $(\mu)^3$He$^+$, the latter closer to unity because the uncertainty due to the helion FFs is the dominant one. The extra inflation factor is $2.42$ and $1.25$ for $(\mu)$H and $(\mu)^3$He$^+$, respectively. Taking into account the additional model dependence of the proton FFs, the FF uncertainty would also dominate in the case of $(\mu)$H.

To control hadronic uncertainties, it is crucial to account for their 
correlations across various correction terms. For example, the recoil 
and Zemach radius effects generate contributions to the HFS of opposite 
sign; consequently, the uncertainty of their sum is smaller than that of 
the pure Zemach radius contribution. Similarly, the $F_2^2(Q^2)$ contribution 
to $W(t)$, discussed below \Eqref{pointlike}, cancels when combined with a 
data-driven dispersive evaluation of the mixed hVP and proton polarizability 
correction.

One notices that our results both for H and $\mu$H, as well as $\mu^3$He$^+$, are at variance with the previous evaluations, while being significantly more precise (where the comparison of uncertainties is possible). To comment on this, we note that the results of \cite{Borie:2011eia} are based on the non-recoil result for $\mu$VP, $n^3 E_{nS\text{-HFS}}^\mathrm{\mu VP}=\nicefrac{3}{4}E_\mathrm{F}\,\alpha(Z\alpha)$, using $E_{nS\text{-HFS}}^\mathrm{hVP}=0.66 \,E_{nS\text{-HFS}}^\mathrm{\mu VP}$ \cite{Borie:1981mr} (it appears that \cite{Borie:2011eia} uses a slightly different prefactor $\simeq 0.667$). While the numerical prefactor underestimates the ratio between the hVP and $\mu$VP contributions, the non-recoil limit without the finite-size effects gives a much bigger integral (as illustrated in Sec.~\ref{sec:recoil_finite_size}). This explains the discrepancy between our results for $\mu$H and $\mu^3$He$^+$ and those of \cite{Borie:2011eia}.

For $\mu$H there is also a calculation of \cite{Faustov:1997rc}, which gives a smaller result than \cite{Borie:2011eia}, although still a bigger one than the result of this work. Here we note that our Eq.\ (\ref{eq:elasticHFS}) reproduces the result of \cite{Faustov:1997rc} with the following caveat: Eq.\ (12) of that reference
appears to contain a mistake, namely, it multiplies the Pauli FF $F_2(Q^2)$ by an extra factor $2$. The final result of \cite{Faustov:1997rc}, the sum of their Eqs.\ (19), (25), and (26), is numerically dominated by the first of these equations, and once the aforementioned extra factor $2$ is removed, their result for $W(t)$ becomes numerically very close to our Eq.\ (\ref{eq:elasticHFS}). However, two further corrections appear to be needed in order to achieve a complete match with our Eq.\ \eqref{eq:elasticHFS}, namely, their Eqs.\ (25) and (26) have to be multiplied by extra factors of $\nicefrac{1}{2}$ and $\nicefrac{M}{2m}$, respectively. This allows us to conclude that \cite{Faustov:1997rc} suffers from calculational errors, leading to incorrect results for the combined hVP--finite-size contribution to the HFS in $\mu$H.

Finally, considering the HFS contribution in the ground state of H and comparing our results with those of \cite{Karshenboim:1996ew}, we see that our result is roughly two times bigger than the `finite size' result of that work. We believe this difference is due to an underestimation of the hVP effects in \cite{Karshenboim:1996ew}, which only takes into account the $\rho$ meson pole contribution in the zero width limit \cite{Karshenboim_1995}.

Next we turn to the Lamb shift in $(\mu)$H and $(\mu)^3$He$^+$, see Table \ref{tab:expcoeff}. The value for the LO hVP contribution to $a_e$ is also provided there to establish a correspondence with the $O(Z^4 \alpha^5)$ limit of the hVP contribution to the Lamb shift, as both are determined by $\Pi'(0)$.
\begin{table*}[t]
    \centering
\caption{The hVP contribution to  
    the Lamb shift in H, ${^3}\mathrm{He}^+$, $\mu$H, and $\mu^{3}$He$^+$. The result for $a_e$ corresponds to Eq.\ \eqref{eq:ae}. Results previously reported in the literature are shown in the top block.  The uncertainty of the pointlike result is solely due to the uncertainty of the employed parametrization of $R(t)$. For the finite-size result in $\mu^3\mathrm{He}^+$, the uncertainties stemming from $R(t)$ and from the elastic FFs are added in quadrature (see text).
    }
    \begin{tabular}{l||S[table-format=4.5,group-minimum-digits=1]|S[table-format=-1.6,group-minimum-digits=1]|S[table-format=-3.7,group-minimum-digits=1]|S[table-format=-3.8,group-minimum-digits=1]|S[table-format=-1.1,group-minimum-digits=1]}
 &\multicolumn{1}{c|}{$a_e\times 10^{14}$}&\multicolumn{1}{c|}{$E_{1S}$(H) [kHz]}&\multicolumn{1}{c|}{$E_{1S}$(${^3}\mathrm{He}^+$) [kHz]}&\multicolumn{1}{c|}{$E_{2S}(\mu\mathrm{H})$ [$\upmu$eV]}&\multicolumn{1}{c}{$E_{2S}(\mu^3\mathrm{He}^+)$ [$\upmu$eV]}\\
    \hline
       \hline
       Previous calculations:    &       &          &            &        \\
  \quad\citet{Karshenboim:2021jsc}          &       &          &            &        \\
  \qquad                           LO hVP, $O(Z^4\alpha^5)$   &189(5)& -3.357(20)&&-11.36(27)  &-224(5) \\
  \qquad                           NLO hVP  &       &-3.401(82)&&-11.43(27)  &-226(5) \\
  \quad\citet{Friar:1998wu}                 &       &-3.40(7)  &&            &        \\
  \quad\citet{Borie:2011eia}                &       &&          &-11(1)      &-221(11)\\  
  \quad\citet{Pachucki:1996zza,Pachucki:1999zza}
                                            &       &&          &-11.3(3)    &        \\  
  \quad\citet{Faustov:1999fg}               &       &&          &-10.949(385)&        \\
  \quad\citet{Martynenko:2001qf}            &       &&          &-10.772(377)&        \\
  \quad\citet{Krutov:2015pxa}               &       &&          &            &-217.0  \\  
        \hline
  This work: &&&& \\
  \quad LO hVP, exact \eref{LOLSexact}&   &-3.388(16) &-54.27(25)&-11.218(53) & -220.3(10) \\
  \quad LO hVP, $O(Z^4\alpha^5)$ \eref{LOLSapprox}
  & 187.5(9) &-3.388(16) &-54.27(25)&-11.260(53) & -222.1(10) \\
\quad hVP--finite-size \eqref{eq:LSweighting} &&0.00008&0.0081&0.055&6.7(2) \\
    \hline
    \end{tabular}
    \label{tab:expcoeff}
\end{table*}
The results for the pointlike contributions agree with the previous calculations within uncertainties. The effect of the higher-order corrections to Eq.\ \eqref{eq:LOLSapprox} is of the order of the uncertainty propagated from $R(t)$ for muonic systems and is completely negligible for electronic systems. The finite-size effects, being $O(Z\alpha)$ corrections, are calculated here for the first time. They are very small for electronic systems, but relevant for muonic systems. While they only approach the size of the  uncertainty of the leading pointlike effect in $\mu$H, the finite-size effect in $\mu^3\mathrm{He}^+$ is about $7$ times bigger than the uncertainty of the pointlike hVP contribution. Note that the ``NLO hVP'' corrections evaluated in \cite{Karshenboim:2021jsc} account for mixed eVP and hVP corrections, as well as lepton vertex corrections with an hVP insertion, and should not be confused with the combined hVP--finite-size correction evaluated in this work.

\section{Summary and Conclusion} \seclab{Sec5}
We have evaluated the hVP contributions to the Lamb shift and the HFS of the $1S$ and $2S$ levels in ordinary and muonic hydrogen atoms and ${^3}$He$^+$ ions, using the DHMZ parametrization of the $R$ ratio~\cite{davier:2019can,Davier:2023fpl}. Throughout, the nuclear recoil and finite-size effects were kept on the same footing as the hVP, and it is their interplay that has shaped the conclusions.

The recoil corrections that are large in Mu~\cite{Sapirstein:1983xr}---where they carry the logarithm of the constituent mass ratio analyzed by Bodwin and Yennie~\cite{Bodwin:1987mj}---are almost entirely suppressed in H, $\mu$H, and $(\mu)^3$He$^+$ by the nuclear elastic FFs, which cut the loop integral off well below the nuclear mass. In this regime, the full weighting function is indistinguishable from the non-recoil case, and recoil ceases to matter. This is the content of the criterion established in Sec.~\ref{sec:recoil_finite_size} for when the recoil corrections to any VP contribution may be neglected (though we explicitly retain them in our evaluation).

Our results for the pointlike hVP contribution to the Lamb shift are consistent with the existing literature within the stated uncertainties \cite{Karshenboim:2021jsc, Borie:2011eia, Pachucki:1996zza}, and have been extended by a first evaluation of the subleading $O(Z^5\alpha^6)$ hVP correction to the finite-size effect. In contrast, the contributions to the HFS show significant deviations from previous evaluations \cite{Karshenboim:1996ew, Faustov:1997rc, Borie:2011eia}. The discrepancies identified in previous HFS evaluations are primarily attributed to two factors. First, some prior estimates relied on a simple rescaling of the $\mu$VP, which works for the Lamb shift but underestimates the ratio between the hVP and $\mu$VP contributions in the HFS. Second, we have identified calculational errors in a previous combined hVP--finite-size evaluation for $\mu$H \cite{Faustov:1997rc}.

These updates carry significant implications for upcoming $\mu$H spectroscopy experiments. In particular, the experimental uncertainty anticipated by the CREMA collaboration for the ground-state HFS in $\mu$H is 1 ppm, or $0.2\ \upmu$eV \cite{Amaro:2021goz}. Our updated results for the hVP contribution in $\mu$H differ from previously published evaluations by approximately ten times this anticipated experimental uncertainty.

\acknowledgments
We thank Fred Jegerlehner for valuable discussions and his help in implementing the required integrals into his \textsc{alphaQED} code.  We thank Misha Eides, Jens Erler, Rodolfo Ferro-Hern\'andez, Misha Gorchtein, and Savely Karshenboim for stimulating discussions, and Marc Vanderhaeghen for useful remarks on the manuscript.
BM acknowledges the fruitful collaborations with Michel Davier, Andreas Hoecker, Anne-Marie Lutz, Andres Pinto, L\'eonard Polat, Zhiqing Zhang, and with the BMW lattice QCD collaboration.

This work is supported by the Deutsche Forschungsgemeinschaft (DFG) 
through the Emmy Noether Programme (grant 449369623) and the Collaborative Research Center 1660 ``Hadrons and Nuclei as Discovery Tools'' (grant 514321794), as well as by the French National Research Agency under contract ANR-22-CE31-0011.

\appendix 
\setcounter{table}{0}
\renewcommand{\thetable}{\Alph{section}\arabic{table}}

\section{More Details on the Hadronic Vacuum Polarization Contributions}
\label{sec:appendixA}
For deeper insights into our calculation we provide the following two tables. Table \ref{tab:expcoeff2} contains the results for the hVP contribution to the ground-state HFS in Mu, $(\mu)$H, and $(\mu)^3\mathrm{He}^+$, corresponding to various limits of $W(t)$ and different inputs for $R(t)$. For the latter, in addition to the DHMZ parametrization, we use Jegerlehner's \textsc{alphaQED23} parametrization \cite{Jegerlehner:2017zsb,Jegerlehner:2019lxt}. The close agreement between the two parametrizations reflects the robustness of these observables against the choice of
$R(t)$ input.  
\begin{table*}[hbt]
    \centering
    \begin{threeparttable}
        \caption{Evaluation of the hVP contributions to the ground-state HFS of Mu, H, ${^3}\mathrm{He}^+$, $\mu$H, and $\mu^{3}$He$^+$, shown here for the various limiting cases discussed in the text.
            \label{tab:expcoeff2}
    }
    \begin{tabular}{l||S[table-format=2.10,group-minimum-digits=1]|S[table-format=2.8,group-minimum-digits=1]|S[table-format=-2.7,group-minimum-digits=1]|S[table-format=2.7,group-minimum-digits=1]|S[table-format=-3.4,group-minimum-digits=1]}
 &\multicolumn{1}{c|}{Mu [kHz]}&\multicolumn{1}{c|}{H [kHz]}&\multicolumn{1}{c|}{$^3$He$^+$ [kHz]}&\multicolumn{1}{c|}{$\mu$H [$\upmu$eV]} &  \multicolumn{1}{c}{$\mu^3$He$^+$ [$\upmu$eV]}\\
    \hline
       \hline
DHMZ &&&&&\\
   \qquad finite size &&&&&\\
      \qquad\qquad full kernel \eref{elasticHFS} &            &0.0860(4)&-0.476(17)&2.153(11)&-15.19(57) \\
   \qquad\qquad non-recoil  \eref{elasticHFSnorecoil} &            &0.0876(4)&-0.483(18)&2.094(11)&-15.24(57)  \\
   \qquad pointlike &&&&&\\
      \qquad\qquad full kernel \eref{pointlike}& 
      &0.2736(12)&-1.860(10)&7.04(3)&-61.0(3)\\
      \qquad structureless &&&&&\\
         \qquad\qquad full kernel \eref{kernelpoint}&  0.2333(11)
         &0.2122(10)&-3.44(2)&5.49(3)&-111.6(5)  \\
   \qquad\qquad non-recoil \eref{pointlikenorecoilHFS}&1.237(6) 
   &0.396(2)&-4.83(2)&9.46(4)&-152.4(7)  \\
   \hline
\textsc{alphaQED23}&&&&&\\  
   \qquad finite size &&&&&\\
      \qquad \qquad full kernel \eref{elasticHFS} &            &0.0860(5) &-0.476(17)&2.153(14)&-15.17(55)  \\
   \qquad \qquad non-recoil \eref{elasticHFSnorecoil}  &            &0.0876(6) &-0.482(18)&2.093(13)&-15.22(56)  \\
   \qquad pointlike &&&&&\\
      \qquad \qquad full kernel \eref{pointlike}  &
      &0.2745(19)&-1.863(12)&7.07(5)&-61.1(4)\\
   \qquad structureless &&&&&\\
   \qquad\qquad full kernel \eref{kernelpoint}& 0.2338(15) &0.2128(14)&-3.45(2) &5.51(4)&-112.0(8)  \\
  \qquad \qquad  non-recoil  \eref{pointlikenorecoilHFS} & 
   1.242(9)
   &0.397(3)&-4.85(3)&9.49(7)&-152.9(11) \\
   \hline
    \hline
    \end{tabular}
    \end{threeparttable}
\end{table*}

Table \ref{tab:expcoeffall} shows the result for the hVP contribution to $a_e$ and the Lamb shift of the ground state in H and ${^3}\mathrm{He}^+$ and $2S$ state in $\mu$H and $\mu^3\mathrm{He}^+$, also obtained using two different inputs for $R(t)$. This table also shows the combined $O(Z^5\alpha^6)$ hVP--finite-size contributions. We omit the uncertainties for the (negligibly small) hVP--finite-size contributions in H and ${^3}$He$^+$, as well as those in $\mu$H (where the hVP--finite-size effect itself is of the size  of the uncertainty of the pointlike hVP contribution).
\begin{table*}[htb]
    \centering
\caption{Evaluation of the hVP contributions to $a_e$, 
    the Lamb shift of H, ${^3}\mathrm{He}^+$, $\mu$H, and $\mu^{3}$He$^+$, shown here for the various limiting cases discussed in the text. 
        \label{tab:expcoeffall}
    }
    \begin{tabular}{l||S[table-format=4.5,group-minimum-digits=1]|S[table-format=-2.7,group-minimum-digits=1]|S[table-format=-3.7,group-minimum-digits=1]|S[table-format=-3.8,group-minimum-digits=1]|S[table-format=-1.1,group-minimum-digits=1]}
 &\multicolumn{1}{c|}{$a_e\times 10^{14}$}&\multicolumn{1}{c|}{$E_{1S}$(H) [kHz]}&\multicolumn{1}{c|}{$E_{1S}(^3\mathrm{He}^+)$ [kHz]}&\multicolumn{1}{c|}{$E_{2S}(\mu\mathrm{H})$ [$\upmu$eV]}&\multicolumn{1}{c}{$E_{2S}(\mu^3\mathrm{He}^+)$ [$\upmu$eV]}\\
       \hline
        \hline
LO hVP, exact \eref{LOLSexact}  && &&&\\  
   \quad DHMZ        &   &-3.388(16)& -54.27(25) &-11.218(53) & -220.3(10)  \\
   \quad\textsc{alphaQED23}
                     &   &-3.381(20)& -54.16(32) &-11.196(66)& -219.9(13) \\
   \hline 
   \hline
 LO hVP, $O(Z^4\al^5)$ \eref{LOLSapprox}  & &&&&\\  
   \quad DHMZ        & 187.5(9) & -3.388(16)&-54.27(25) &-11.260(53) & -222.1(10) \\
   \quad\textsc{alphaQED23}
                     & 187.1(11) & -3.381(20)&-54.16(32) &-11.238(66)& -221.7(13)\\   \hline 
    \hline
hVP--finite-size & & & & & \\
\quad
DHMZ  & &&&&\\  
   \qquad full kernel, finite size \eref{LSweighting}&&0.00008&0.0081&0.055&6.7(2)\\
      \qquad non-recoil, finite size \eref{LSweightingnonrecoil}  & &0.00009&0.0083&0.055&6.8(2)\\
   \qquad pointlike \eref{LSweightingpointlike}  & &0.00003&0.0001&0.019&0.1\\
\quad
\textsc{alphaQED23}  && &&&\\  
   \qquad full kernel, finite size \eref{LSweighting}&&0.00008&0.0081&0.055&6.7(2)\\
      \qquad non-recoil, finite size \eref{LSweightingnonrecoil}  & &0.00009&0.0083&0.055&6.8(2)\\
   \qquad pointlike \eref{LSweightingpointlike}  & &0.00003&0.0001&0.019&0.1\\
            \hline    
    \end{tabular}
\end{table*}

\clearpage

\end{document}